\documentclass[conference]{IEEEtran}
\usepackage{url}
\usepackage{graphicx}
\usepackage{amsmath}
\usepackage{booktabs}
\usepackage{multirow}
\begin{document}
\title{Machine Learning-Driven Anomaly Detection for 5G O-RAN Performance Metrics}
\author{\author{Babak Azkaei, Kishor Chandra Joshi, George Exarchakos\\
Email: \{b.azkaei, k.c.joshi, g.exarchakos\}@tue.nl
\thanks{....}}}
\markboth{Eindhoven University of Technology}
{Shell \MakeLowercase{\textit{et al.}}: Bare Demo of IEEEtran.cls for IEEE Journals}
\maketitle
\begin{abstract}
The ever-increasing reliance of critical services on network infrastructure coupled with the increased operational complexity of beyond-5G/6G networks necessitate the need for proactive and automated network fault management. The provision for open interfaces among different radio access network\,(RAN) elements and the integration of AI/ML into network architecture enabled by the Open RAN\,(O-RAN) specifications  bring new possibilities for active network health monitoring and  anomaly detection. In this paper we leverage these advantages  and develop an anomaly detection framework that proactively detect the possible throughput drops for a UE and minimize the post-handover failures. We propose two actionable anomaly detection algorithms tailored for real-world deployment.
The first algorithm identifies user equipment (UE) at risk of severe throughput degradation by analyzing key performance indicators (KPIs) such as resource block utilization and signal quality metrics, enabling proactive handover initiation. The second algorithm evaluates neighbor cell radio coverage quality, filtering out cells with anomalous signal strength or interference levels. This reduces candidate targets for handover by 41.27\% on average. Together, these methods mitigate post-handover failures and throughput drops while operating much faster than the near-real-time latency constraints. This paves the way for self-healing 6G networks.
\end{abstract}
\IEEEpeerreviewmaketitle
\section{Introduction}
\IEEEPARstart{T}{he} evolution from 5G to 6G promises significant advancements in service quality, including enhanced data rates, reduced latency, and improved connectivity, fostering innovations in Internet of Things (IoT), autonomous systems, and immersive experiences \cite{R01}. A key enabler of this evolution is Open Radio Access Networks (O-RAN), which offers a disaggregated architecture with standardized open interfaces, cloudification, programmability, and automation driven by artificial intelligence (AI) \cite{R02}. These principles aim to make 6G networks more agile, cost-effective, energy-efficient, and resilient. However, the increased complexity of 6G networks poses challenges in maintaining reliability and performance. Traditional fault detection mechanisms, relying on static thresholds and rule-based systems, often fall short in addressing the dynamic and unpredictable behavior of wireless environments. Ensuring robust and sustainable service delivery for mission-critical use cases such as remote surgery and autonomous driving requires advanced fault management approaches.

To address the above challenges, fault management must evolve from predefined methods to proactive anomaly detection. By leveraging machine learning (ML) and statistical methods, anomaly detection can identify subtle deviations in network key performance indicators (KPIs) such as spikes in latency, channel quality degradation, irregular resource utilization, or low throughput that often precede faults. Early detection allows for timely intervention, preventing minor issues from escalating into major failures and maintaining network resilience. One of the key features of the O-RAN architecture is the ability to integrate AI/ML models to detect any abnormal network behavior by observing the collected performance metrics.

Anomaly detection provides valuable insights into network performance, enabling proactive responses that minimize disruptions, ensure high availability, and support self-healing. Effective network fault management benefits from a hierarchical approach: near-real-time detection enables immediate corrective actions, while non-real-time analysis diagnoses root causes for long-term optimization. This dual-layer strategy ensures both rapid response to critical issues and sustained network resilience.
This integration of AI/ML and RAN components enabled by O-RAN architecture has received considerable attention from research community in recent years demonstrating the potential of anomaly detection for a resilient network operation. In \cite{nediyanchath2020anomaly}, a multi-scale convolutional recurrent encoder-decoder network is proposed, demonstrating effectiveness in detecting anomalous LTE network windows. The study focuses on windows rather than individual KPIs. \cite{yuan2020anomaly} introduced an AI-based anomaly detection and root cause analysis (RCA) method with low computational complexity, capable of identifying performance degradation in large-scale networks. However, this research lacks a detailed discussion of inference time. In \cite{sundqvist2022uncovering}, a combined learner approach is proposed for detecting latency anomalies. While effective for latency detection, this study does not address throughput degradation or provide model interpretability. Furthermore, \cite{mahrez2023benchmarking} conducted a benchmarking study on various anomaly detection algorithms to optimize handovers in O-RAN environments. However, this work does not differentiate between serving cell and neighbor cell KPIs, nor does it incorporate common explainable AI (XAI) methods. Lastly, \cite{sun2024spotlight} proposed SpotLight, a framework that is able to continuously detect and localize anomalous KPIs from both RAN and platform layer. Due to the aggregate nature of the data collected, complex anomaly explanation phase and observation window of 6.4s, SpotLight might not be useful for time-sensitive recovery mechanisms like handover.

Many AI/ML models, although powerful, often lack explainability on their own, making it necessary to augment explanation to these models. Explainable AI is an emerging area that enhances the transparency and interpretability of AI systems, ensuring that humans understand the reasons behind AI-driven decisions.  Primarily there are two types of XAI methods: (i) Model-specific methods that are designed for specific types of model; e.g. permutation feature importance for decision trees and (ii) Model-agnostic methods that can be applied to any ML model without dependency on its architecture. In case of model-agnostic methods, techniques like SHapley Additive exPlanations (SHAP) \cite{NIPS2017_7062} and Local Interpretable Model-agnostic Explanations (LIME) create interpretable estimations of the model’s predictions. XAI techniques can provide insight into why specific data points are flagged as anomalies, and what features contributed the most to a detected anomaly. This can help foster trust and facilitate more effective interventions and RCA.

Motivated by the mentioned requirements and limitations of the state-of-the-art techniques, in this work, we introduce an anomaly detection framework. Our framework enhances the user throughput and the handover experience utilizing the RAN KPIs. Our key contributions are summarized as follows:
\begin{itemize}
\item We propose two anomaly detection algorithms that predict user equipment (UE) throughput degradation and also flag neighbor cells with poor coverage to enhance handover reliability.
\item We demonstrate that our fine-tuned ML algorithms execution time is faster than the O-RAN near-RT RIC time constraints and thus capable of real-time operation.
\item We augment anomaly detection models with XAI methods. We observed that these explanations provided accurate insight about KPIs having the most impact on anomalous behavior.
\item Both algorithms are highly accurate, actionable and can directly work with handover or other recovery mechanisms.
\end{itemize}
The rest of the paper is structured as follows. Section II details our system model and methodology, defining critical KPIs, anomaly types, and ML models. Section III presents ML model performance and latency; and explainability analysis of the ML models. Section IV discusses the outcomes of this research and concludes with future directions.

\section{System model and Methodology}
\begin{figure}[t!]
\centering
\includegraphics[scale=0.31]{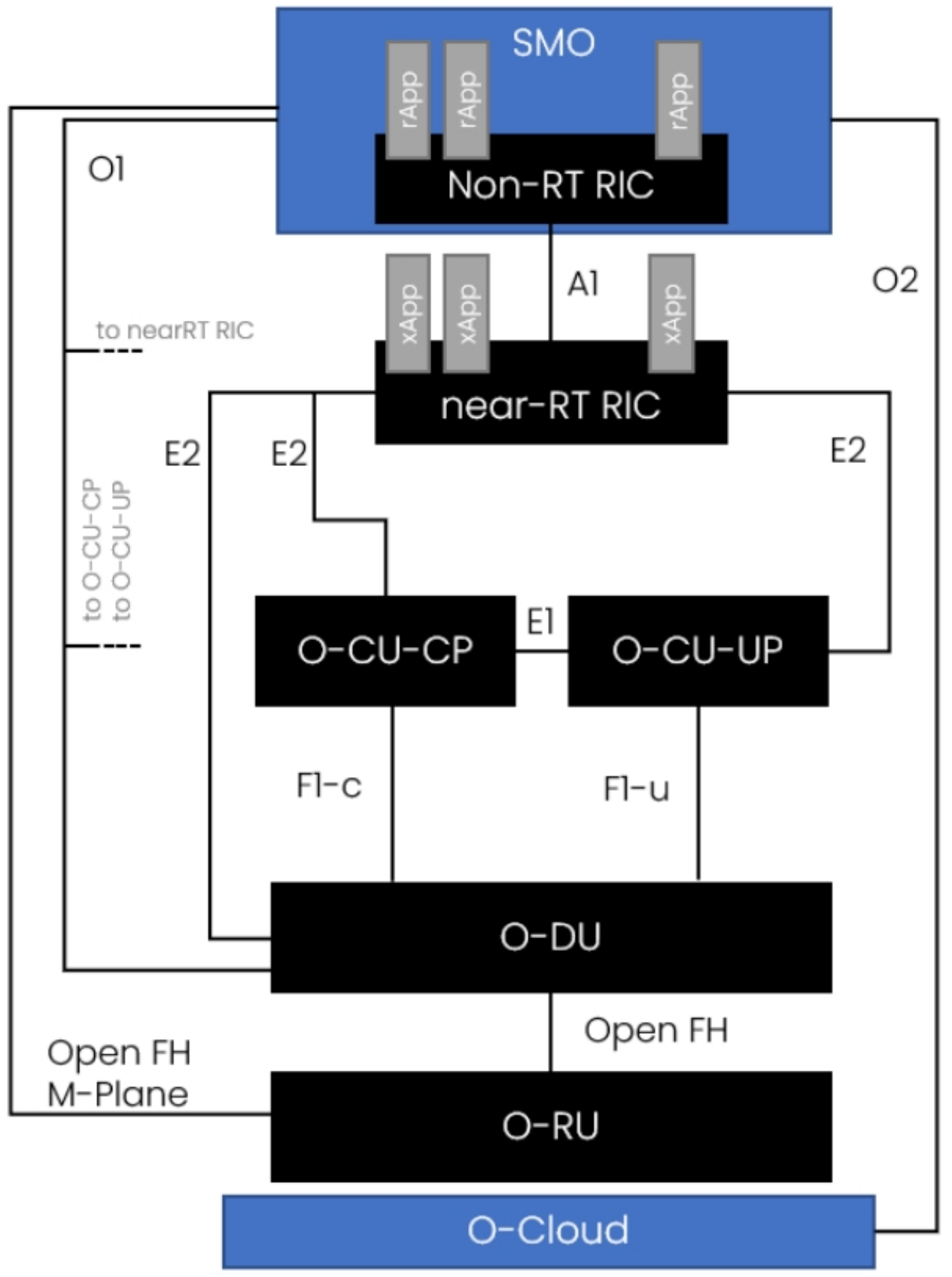}
\caption{O-RAN Architecture \cite{kliks2023towards}.}
\label{F_AO}
\end{figure}

We consider the O-RAN architecture shown in Figure \ref{F_AO}. The key elements of O-RAN architecture include near-real-time and non-real-time RAN Intelligent Controllers (RICs), open interfaces (E2, A1, O1, O2, F1, etc.) and a disaggregated deployment of radio unit (RU), distributed unit (DU) and the centralized unit (CU) following the 7.2x split of the 3GPP 5G specifications. The most relevant interfaces for our research in this paper are the E2 and A1 interfaces. E2 is a logical interface connecting the near-RT RIC with an E2 node (RU, DU and CU) and is used for collecting the information (metrics) related to UEs, cells and slices, etc. The A1 interface connects non-RT-RIC and near-RT-RIC and enables the management of ML services and policies. The E2 Service Model (E2SM) has three different categories out of which E2SM Key Performance Measurement (KPM) \cite{R13} is the most relevant E2 service model for anomaly detection tasks. KPM focuses on tracking real-time network metrics from different network entities, including UEs, and anomaly detection identifies deviations from normal behavior in these metrics, allowing operators to act before problems escalate into service-impacting events. By augmenting anomaly detection to the KPM aspect of the E2 service model, operators can achieve a smarter and more robust RAN management system.

\begin{figure}[t!]
\centering
\includegraphics[scale=0.8]{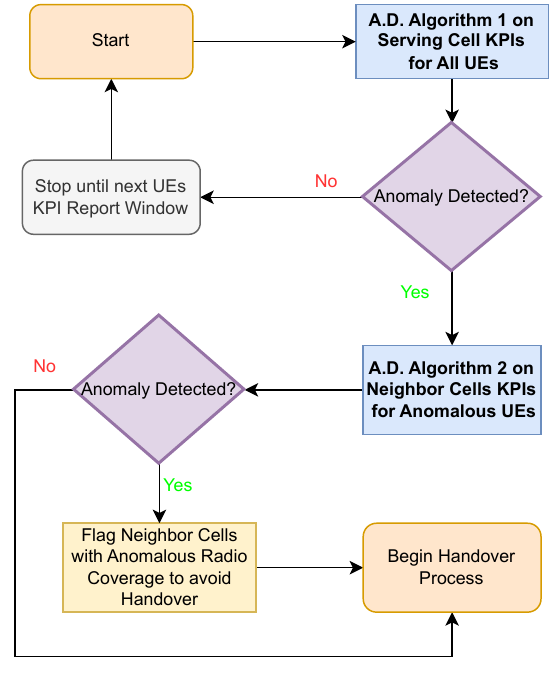}
\caption{Summary of our performance anomaly detection approach.}
\label{Flowchart}
\end{figure}

We focus on two key areas as part of 5G O-RAN fault management: (1) detecting serving cell anomalies that predict UE throughput degradation, and (2) identifying neighbor cells with suboptimal coverage to prevent handover failures. Our approach involves two anomaly detection models, summarized in Figure \ref{Flowchart}:
\begin{enumerate}
\item Serving Cell Anomaly Detection: Detects UEs with anomalous serving cell KPIs that result in low throughput. These anomalies trigger recovery mechanisms, such as handovers.
\item Neighbor Cell Anomaly Detection: Identifies anomalous neighbor cell KPIs to reduce candidate cells for handover, ensuring reliable connectivity.
\end{enumerate}
Our anomaly detection framework aligns with the O-RAN architecture, and leverages the capabilities of RIC in the following way:
\begin{itemize}
\item Near-real-time RIC: Hosts the anomaly detection algorithm as an xApp, leveraging the E2 KPM service model for closed-loop control (10ms to 1s). Detected anomalies trigger immediate corrective actions.
\item Non-real-time RIC: Manages training, model updates, and explanations as an rApp with a control timescale exceeding 1 second. This layer can perform deeper analysis for root cause identification and policy optimization.
\end{itemize}
\subsection{KPIs and Data Collection Flow}
The anomaly detection pipeline relies on KPIs related to resource utilization and signal quality. The KPIs we use are described as below:
\begin{itemize}
\item Physical Resource Blocks (PRB) Usage for Downlink: Indicates the utilization of PRBs in downlink for UE.
$\text{PRB}_{\text{max}} = \frac{\text{Total Bandwidth (Hz)}}{\text{Subcarrier spacing (Hz)} \times {\text{Subcarriers per PRB}}}$
\(\text{Total Bandwidth} = 100 \, \text{MHz}, \quad \text{Subcarrier spacing} = 30 \, \text{kHz}, \quad \text{Subcarriers per PRB} = 12.\)
Thus, maximum number of PRBs in our setting is 273.
\item Reference Signal Received Power (RSRP): Measures the average power of resource elements carrying reference signals. $\mathrm{RSRP}=\frac{1}{N} \sum_{i=1}^N P_i$, where $P_i$ is the power of the $\mathrm{i}th$ resource element that carries the reference signal. $\mathrm{N}$ is the number of resource elements containing the reference signal.
\item Received Signal Strength Indicator-to-Interference plus Noise Ratio (RSSINR) $=\frac{RSSI}{P_{\text {interference }}+P_{\text {noise }}}$, where $RSSI$ is Received Signal Strength indicator, representing the total received power (including signal, interference, and noise) in the specified bandwidth. $P_{\text {interference }}$ is the total power of all interfering signals and $P_{\text {noise }}$ is the noise power.
\item Reference Signal Received Quality (RSRQ) $=\frac{N \cdot \mathrm{RSRP}}{\mathrm{RSSI}}$, where $\mathrm{N}$ is the number of resource blocks over which RSSI is measured.
\end{itemize}
To contextualize how UEs data is collected and processed:
\begin{itemize}
\item UEs periodically transmit radio frequency (RF) measurements (e.g., RSRP, RSSINR) and other relevant KPIs to the O-RU over the 5G air interface, providing critical data on signal quality and quality of service (QoS).
\item The O-RU processes these RF measurements and generates IQ (In-phase/Quadrature) samples, which are sent to the O-DU via the Open Fronthaul interface for further processing.
\item The O-DU aggregates KPIs (e.g., PRB utilization, RSRQ, throughput) from connected UEs. The data is tagged with its DU ID for traceability and packaged for upstream analysis.
\item The O-DU streams the tagged KPI data to the Near-RT RIC over the E2 interface, where it is stored in a database. Anomaly detection algorithms, powered by ML models trained offline by the Non-RT RIC and deployed via the Service Management and Orchestration (SMO) framework, analyze the data in real time to identify anomalies.
\end{itemize}

\subsection{Problem Definition}
The objective is to develop lightweight models to predict throughput degradation and handover issues caused by:
\begin{itemize}
\item Type 1 Anomalies: PRB contention in serving cells, identified via downlink PRB usage.
\item Type 2 Anomalies: Radio coverage issues in serving and neighbor cells, detected via RSRP, RSSINR, and RSRQ.
\end{itemize}
An instance is labeled as anomalous ($y = 1$) if the observed throughput $T_\text{obs}$ is significantly lower than the target throughput $T_\text{target}$. We can predict low throughput for those UEs, with the accuracy of the ML model directly influencing the reliability of these predictions. Anomalies trigger recovery mechanisms like handover.

Missing or inconsistent KPI reports from UEs can skew anomaly detection, especially with long reporting intervals (low temporal granularity). To ensure uniform reporting across UEs, missing KPI values may require interpolation, which ensures consistency but can reduce data accuracy. We prioritized raw, non-interpolated data to maintain more accurate anomaly detection. For short reporting intervals, a potential solution is to define a time window that flags a UE as anomalous when it exceeds a threshold number of anomalies within that period. This approach capture cumulative irregularities while minimizing false positives.
\subsection{Anomaly Detection Models}
This section outlines the machine learning models used for anomaly detection in our study, categorized by their learning paradigm and operational characteristics.
\begin{itemize}  
\item Isolation Forest (Unsupervised):  It constructs an ensemble of binary trees to isolate anomalies through recursive random partitioning of feature space. It has a linear time complexity ($O(n)$) and thus efficient for high-dimensional data.  However, it has limited sensitivity to complex feature interactions and local anomalies.  
\item Random Forest (Supervised): it consist of ensemble of decision trees trained on bootstrapped samples, with anomaly scores derived from majority voting or class probability thresholds. It is robust to overfitting and handles class imbalance via bagging and feature randomness. However, performance degrades with noisy labels.  
\item AutoEncoder (Unsupervised): This employs neural networks with a bottleneck architecture that learns compressed latent representations. Anomalies are flagged via reconstruction error ($||x - \hat{x}||^2$). It is able to captures non-linear relationships in high-dimensional KPI streams. It is computationally intensive to train and sensitive to hyperparameter choices.    
\item AutoEncoder-1SVM (Hybrid) \cite{nguyen2019scalable}: It uses a two-stage pipeline: (1) AutoEncoder reduces dimensionality, (2) One-Class SVM (Support Vector Machine) separates anomalies in latent space using a kernelized hypersphere. It combines feature abstraction (AutoEncoder) with kernel-based outlier sensitivity (SVM). The key limitations are increased latency from sequential training and requires tuning two models.   
\end{itemize}
\subsection{ML Models Performance Metrics}
We compare ML models for anomaly detection using the metrics as follows. True Positives~(TP) is the number of correctly predicted positive cases while False Positives (~FP) is the number of negative cases incorrectly flagged as positive. False Negatives~(FN) is the number of positive cases incorrectly flagged and True Negatives~(TN) is the correctly predicted negative cases. We define\,$\text{Precision} = \frac{\text{TP}}{\text{TP} + \text{FP}}$,  and $\text{Recall} = \frac{\text{TP}}{\text{TP} + \text{FN}}$.
\newline
F1-Score which balances precision and recall using their harmonic mean can be defined as $\text{F1-Score} = 2 \cdot \frac{\text{Precision} \cdot \text{Recall}}{\text{Precision} + \text{Recall}}$
\newline
 Finally, the model accuracy is defined as  $\text{Accuracy} = \frac{\text{TP} + \text{TN}}{\text{TP} + \text{TN} + \text{FP} + \text{FN}}$

\section{Simulation Results}
For our simulations, we used a dataset from the O-RAN Software Community available on GitHub \cite{rdataset}. This dataset contains 10,000 KPI reports from 20 users with different mobility patterns in a 5G Network. Each row contains the following information:
\begin{itemize}
\item PRB Used for Downlink (serving cell)
\item Radio Coverage metrics (RSRP, RSSINR, RSRQ) from serving and neighbor cells
\item Data Radio Bearer (DRB) UE Throughput Downlink, Target (desired) Throughput
\item Other information including DU, Cells, and UE ID, etc.
\end{itemize}
We utilized 4 KPIs from the serving cell and 15 KPIs from five neighbor cells, resulting in a total of 19 features as input to our ML models. We implemented our solution using Python, with the following open source libraries being the most crucial:
\begin{itemize}
\item Scikit-learn for Isolation Forest and Random Forest models, hyperparameters tuning and permutation importance.
\item PyOD \cite{chen2024pyod}, and PyTorch for training AutoEncoder models. 
\item SHAP for model explanation.
\end{itemize}
\subsection{Anomaly Detection Algorithm 1}
Table \ref{tab_parameters} provides a summary of key hyperparameters tuned during the training phase. Due to space constraints, some hyperparameters are not listed.
For hyperparameter tuning, multiple configurations of each model were evaluated based on their F1-scores, calculated using anomaly labels from the training data. The model configuration with the highest F1-Score was selected. It is critical to perform hyperparameter tuning exclusively on the training data to prevent data leakage and mitigate the risk of overfitting.
\begin{table}[ht!]
\centering
\caption{Important Hyperparameters of Evaluated Models}
\begin{tabular}{@{}lcl@{}}
\toprule
Model & Parameter & Value \\
\midrule
\multirow{2}{*}{Isolation Forest}
& max\_samples & 0.005 \\
& n\_estimators & 200 \\
\midrule
\multirow{3}{*}{Random Forest} 
& min\_samples\_leaf & 1 \\
& min\_samples\_split & 2 \\
& n\_estimators & 200 \\
\midrule
\multirow{5}{*}{AutoEncoder} & batch\_size & 16 \\
& dropout\_rate & 0.05 \\
& epoch\_num & 100 \\
& hidden\_activation\ & relu \\
& hidden\_neurons\ & [32, 16, 16, 32] \\
\midrule
\multirow{5}{*}{AutoEncoder-1SVM} & batch\_size & 16 \\
& dropout\_rate & 0.3 \\
& epoch\_num & 75 \\
& hidden\_activation & tanh \\
& hidden\_neurons & [16, 8, 8, 16] \\
& Sigma ($\sigma$) & 1 \\
& Nu ($\nu$) & 0.1 \\
\bottomrule
\end{tabular}
\label{tab_parameters}
\end{table}
\newline
Table~\ref{tab_performance} compares the performance of four anomaly detection models on unseen test data, evaluated using metrics discussed before. We included Precision and Recall metrics for anomalies and F1-Score is macro-averaged.
\begin{table}[ht!]
\centering
\caption{Performance Metrics of Models on Test Data}
\begin{tabular}{@{}lcccc@{}}
\toprule
Model & Precision (1) & Recall (1) & F1-Score & Accuracy \\
\midrule
Isolation Forest & 0.95 & 0.51 & 0.79 & 87\% \\
Random Forest & 0.86 & 0.86 & 0.90 & 93\% \\
AutoEncoder & 0.78 & 0.73 & 0.84 & 88\% \\
AE-1SVM & 0.83 & 0.68 & 0.84 & 88\% \\
\bottomrule
\end{tabular}
\label{tab_performance}
\end{table}
\subsubsection{Key Observations}
\begin{itemize}
\item Random Forest dominance: The Random Forest model achieves the highest F1-score (0.90) and accuracy (93\%), demonstrating its robustness in balancing precision and recall. This suggests a strong generalization for supervised scenarios with labeled anomalies in offline training. For our use case, this translates to highly accurate handover alerts and a reduced incidence of false or unnecessary handovers.
\item Isolation Forest trade-offs: While Isolation Forest has the highest precision (0.95), its low recall (0.51) indicates frequent missed anomalies, likely due to its unsupervised partitioning strategy. The resultant F1-score (0.79) limits its utility in recall-sensitive deployments.
\item AutoEncoder vs. AE-1SVM: The standalone AutoEncoder achieves higher recall (0.73 vs. 0.68) than the hybrid AE-1SVM, but the latter improves precision (0.83 vs. 0.78). This reflects the AutoEncoder’s sensitivity to subtle anomalies versus the AE-1SVM’s reliance on One-Class SVM’s tighter decision boundaries.
\end{itemize}
\subsubsection{Practical Implications}
\begin{figure}[ht]
\centering
\includegraphics[scale=0.64]{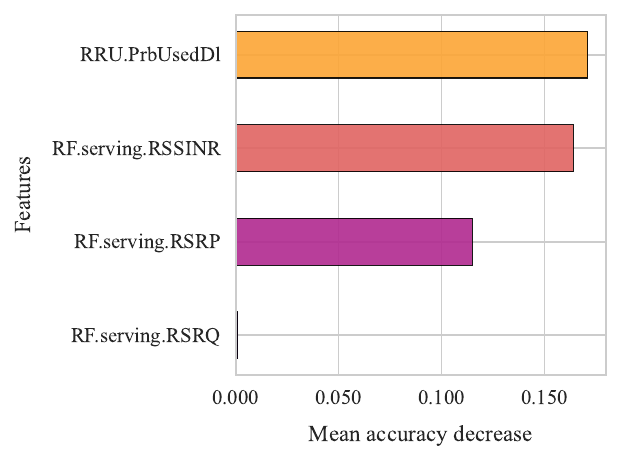}
\caption{Permutation Feature Importance For Random Forest Model.}
\label{F_PME}
\end{figure}

\begin{figure}[ht]
\centering
\includegraphics[scale=0.64]{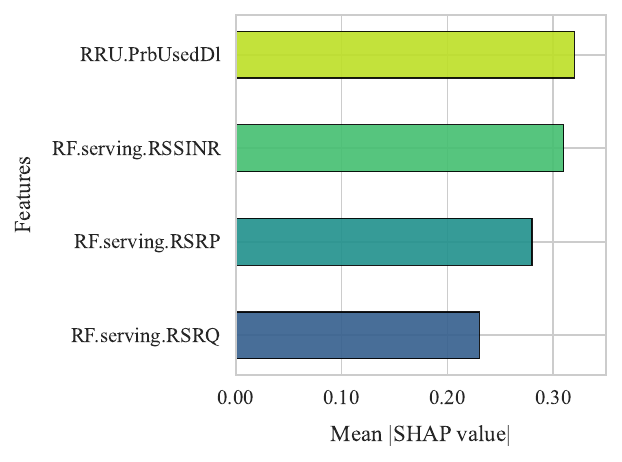}
\caption{SHAP For Isolation Forest Model Explanation.}
\label{F_SHAP}
\end{figure}

Table~\ref{tab_inference} presents inference latency measurements, which are critical for real-time O-RAN applications. Inference latency is measured under identical settings and hardware configurations for all models.
\begin{table}[ht!]
\centering
\caption{Inference Time of Models for 20 Users (ms)}
\begin{tabular}{@{}lc@{}}
\toprule
Model & Average Inference Time (ms) \\
\midrule
Isolation Forest & $0.1926$ \\
Random Forest & $0.2240$ \\
AutoEncoder & $1.944$ \\
AutoEncoder-1SVM & $2.487$ \\
\bottomrule
\end{tabular}
\label{tab_inference}
\end{table}
\begin{itemize} 
\item Latency Compliance: 
\newline The measured inference times of ML models are significantly below the Near‑RT RIC control loop time scale. The fastest model, Isolation Forest, requires only 0.19\,ms per inference for all 20 UEs (one KPI report per UE) while even the slowest model, AE‑1SVM, completes inference in just 2.49\,ms. These times fall comfortably within the Near‑RT constraints, ensuring that anomaly alerts for handover decisions can be generated in a timely manner. These numbers represent only the ML inference times; additional latency will be incurred by data collection and real‑time processing of KPI reports within O‑RAN architecture. Nonetheless, these results demonstrate the potential for implementing real‑time anomaly detection solution. Furthermore, since ML inference typically processes each sample (or UE KPI report) independently, the total inference time scales linearly with the number of UEs (i.e., the total number of KPI reports in a given time window).
\item Model Selection Trade-offs:
\newline Isolation Forest: Its speed (0.19\,ms inference) is unmatched, but a low recall (0.51) may miss subtle anomalies, potentially delaying handovers.
\newline Random Forest: Delivers high reliability with an F1-score of 0.90, making it an excellent choice for detecting anomalies critical to triggering handovers.
\newline AutoEncoder: Offers higher recall (0.73), reducing missed anomalies and enhancing handover robustness, albeit with a moderate latency of 1.92\,ms.
\newline AE-1SVM: High precision (0.83) minimizes false positives, preventing unnecessary handovers that could impact UE throughput. However, its hybrid architecture, despite fewer hidden neurons, results in slightly slower performance compared to the standalone AutoEncoder.
\end{itemize}
The choice of ML model depends on the application requirements. Random Forest offers a balanced trade-off between performance and inference time, making it well suited for this kind of anomaly detection in 5G O-RAN, especially in scenarios with a high number of UEs. AutoEncoders, despite their slightly higher latency, may be preferred for use cases with a larger number of KPI features.
\newline
Figure \ref{F_PME} shows Random Forest model explanation based on permutation feature importance (model-specific) method. Mean accuracy decrease, measures how much a model’s accuracy drops when the values of a particular feature are randomly shuffled.
\newline
Figure \ref{F_SHAP}, shows Isolation Forest explanation based on average absolute SHAP values which is a model-agnostic explanation method.
Permutation Feature Importance and SHAP analysis reveal PRB utilization and RSSINR as the key drivers of our performance anomaly detection in 5G O-RAN. Permutation importance highlights PRB’s role in model accuracy while showing RSRQ’s minimal impact when randomized. SHAP values confirm PRB’s direct influence on anomaly predictions and suggest RSRQ has moderate, context-dependent relevance. This contrast indicates that while PRB utilization and RSSINR are strong, consistent performance indicators, RSRQ’s importance may depend on non-linear effects or localized anomalies.
\subsection{Anomaly Detection Algorithm 2}
Building upon the serving cell KPI-based anomaly detection method introduced earlier, we now propose a strategy to preemptively avoid handovers to neighbor cells with poor radio conditions. While Algorithm 1 flags UEs at risk of throughput degradation based on serving cell KPIs, we extend this approach by detecting anomalous radio coverage in neighbor cells ~(handover candidates) to mitigate the risk of handover failures and post-handover throughput drops.
\begin{figure}[t!]
\centering
\includegraphics[scale=0.84]{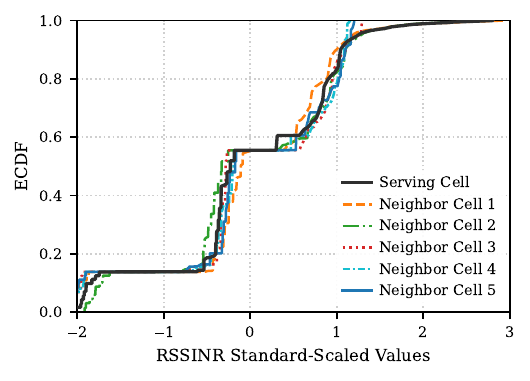}
\caption{Comparison of RSSINR distribution among cells.}
\label{F_CDF}
\end{figure}
Figure~\ref{F_CDF} reveals minimal divergence between the ECDF (Empirical Cumulative Distribution Function) of RSSINR values for serving and neighbor cells, including anomalous cases. Similar patterns were observed for RSRP and RSRQ metrics. To train the second model, we used a modified dataset derived from the previous analysis, excluding samples where PRB contention exceeded 70\% utilization, as these cases were identified as contributing to anomalies. This model was utilized to detect poor radio coverage from neighbor cells. Simulation results revealed that, on average, \textbf{41.27\%} of the neighbor cell KPI reports were identified as anomalous, roughly two out of five cells, from the UE's perspective. This outcome is beneficial because it narrows the selection to one of the three remaining neighbor cells for a more reliable handover. Furthermore, the average inference time for 20 UEs is approximately \textbf{1.7 milliseconds}, demonstrating the model's efficiency for real-time predictions.
Algorithm 2 exhibits significantly higher latency compared to Algorithm 1 when employing the same model type (Random Forest), which is justified by the larger number of input features derived from the five neighbor cells.
To further enhance its functionality, this algorithm can be integrated with a QoS (Quality of Service) prediction model to rank the three remaining cells based on throughput predictions, facilitating an optimal handover decision. Second anomaly detection algorithm provides a less complex input for recovery stage, minimizes computational overhead and simultaneously reduces the risk of handover failures caused by poor radio coverage. Together, Algorithms 1 and 2 complement each other by providing the necessary information for handover process and improving its reliability, respectively. Handover mechanism details remain outside the scope of this work.
\section{Conclusion and Future Work}
This study explored ML-driven performance anomaly detection for 5G O-RAN architectures. By analyzing resource utilization and radio signal quality indicators, it demonstrated how data-driven models can enhance network reliability and preemptively mitigate issues such as throughput degradation and also handover failures. The findings highlight the potential of lightweight, efficient anomaly detection frameworks to operate within stringent near-real-time constraints, offering actionable and explainable insights for current 5G deployments and laying the groundwork for future 6G systems.
\newline Looking ahead, several promising directions emerge. A key next step is validating the proposed approach on a live O-RAN testbed. In addition, expanding anomaly detection to incorporate cross-domain metrics across RAN, cloud infrastructure, and end-user devices could enable comprehensive network performance monitoring. Furthermore, advances in generative AI, such as large language models (LLMs), present opportunities to automate diagnostic reporting and improve operator decision-making through intuitive, narrative-driven insights.
\section*{Acknowledgment}
This work was supported by the European Union’s Horizon Europe Marie Skłodowska-Curie Action SCION under No. 101072375.
\bibliography{references}
\bibliographystyle{unsrt}
\end{document}